# Comment on "Raman spectra of misoriented bilayer graphene"

Zhenhua Ni, Yingying Wang, Ting Yu, Yumeng You, and Zexiang Shen[*]

*Division of Physics and Applied Physics, School of Physical and Mathematical Sciences, Nanyang Technological University, Singapore 637371, Singapore*

**Abstract**

In a recent paper [Phys. Rev. B **78**, 113407 (2008)], Poncharal et al. studied the Raman spectra of misoriented bilayer graphene. They found that the blueshift of 2D band of misoriented graphene relative to that of single layer graphene shows a strong dependence on the excitation laser energy. The blueshift increases with decreasing excitation energy. This finding contradicts our explanation of reduction of Fermi velocity of folded/misoriented graphene [Ni et al. Phys. Rev. B **77**, 235403 (2008)]. In this comment, we present more experimental results from our group as well as from others to show that the blueshift is indeed only weakly dependent on excitation energy. We therefore suggest that our explanation of 2D blushift of folded graphene due to reduction of Fermi velocity is still valid.

The electronic properties of graphene are highly sensitive to the number of graphene layers and also the stacking geometry.[1] Recent studies on misoriented multi-layer epitaxial graphene on SiC substrate revealed the two dimensional Dirac-like character of electronic states.[2,3] Theoretical calculations of electronic structure of bilayer graphene (BLG) with misoriented second layer were also carried out [4,5] and showed that the low energy dispersion of twisted two-layer graphene is linear, similar to that in SLG. Raman spectra of graphene are very sensitive to the electronic band structure of graphene and hence can be used to investigate the electronic structure of BLG that deviates from the AB stacking.

In a recent publication, Poncharal et al.[6] studied the Raman spectra of bilayer graphene in which the two layers are arbitrarily misoriented. They observed a blueshift of 2D band frequency of misoriented bilayer graphene relative to that of single layer graphene (SLG). This result is similar to our observation on folded graphene, where the two SLGs stack with arbitrary orientation.[7] However, Poncharal et al. observed that the blueshift of 2D band increases with decrease in excitation laser energy, which is 2 $cm^{-1}$, 6 $cm^{-1}$ and 9 $cm^{-1}$ for 2.41 eV, 2.33 eV and 1.96 eV excitations, respectively, as shown by the green triangles in Figure 1. This finding contradicts our explanation of reduction of Fermi velocity of folded graphene.[7] They therefore concluded that the blueshift of folded graphene is unlikely to be due to the electronic band structure change, and they attributed it as due to change in phonon dispersion.

We have also measured Raman spectra of folded/misoriented graphene samples with different excitation energy and found that the blueshift is only weakly dependent on excitation energy, which is very different from the results observed by Poncharal et al. The optical imaging of folded graphene sample with size of ~20 um$^2$ is shown in the inset of Fig 1. Both the folded graphene and nearby SLG come from the same piece of SLG, which indicates that the blueshift of 2D band folded graphene is purely due to folding, and not because of the difference in SLG. We have also carried out Raman imaging on the folded graphene with a scanning step of ~100 nm. The folded graphene can be precisely located in the Raman imaging and the average Raman frequency from the folded area is also obtained (Ref. 7, Fig. 1).[7] The four samples we tested give similar results: the 2D blueshift of folded sample is only weakly dependent on excitation energy as shown in Figure 1. Our results are in line with the results of P. C. Eklund's group in Pennsylvania State University. Their results also showed that the blueshift of 2D frequency of folded/misoriented graphene is similar with different excitation energy, as shown by the red squares in Figure 1.[8] The much different observation by Poncharal et al., i.e. blueshift of 2D band largely increases with decrease in excitation laser energy, might be only valid for very special folded/misoriented sample. Based on above results, we suggest that it is premature for Poncharal et al. to draw conclusion that the blueshift is not related to reduction of Fermi velocity.

Finally, we would like to mention that recent scanning tunneling microscopy (STM) study on graphene placed on graphite surface has also revealed the reduction of Fermi velocity of such misoriented sample, where the reduction is ~5% in high energy regime and ~25% in low energy regime.[9] As such system is similar to the folded graphene configuration, their results support our explanation of reduction of Fermi velocity on folded graphene.[7]

In summary, we report our and other group's results on the excitation energy dependent blueshift of 2D band of folded graphene. The results show that the blueshift of folded graphene is only weakly dependent on excitation energy. We suggest that the explanation of 2D blushift of folded graphene due to reduction of Fermi velocity is still valid.

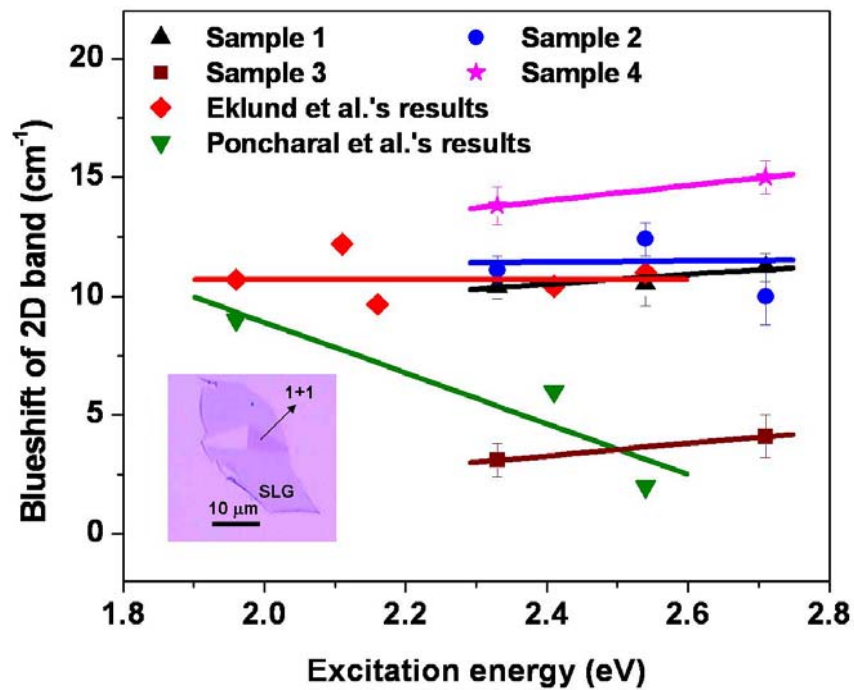

Figure 1 The blueshift of 2D band of folded graphene relative to SLG at different excitation energy. The results from P. C. Eklund's group[8] and Poncharal et al.[6] are also included. The lines are linear fit of the data. The inset is an optical imaging shows our SLG sample and also the 1+1 layer folded graphene. The size of folded area is ~20 μm$^2$.